\documentclass{article}

\usepackage{PRIMEarxiv}

\usepackage[utf8]{inputenc} % allow utf-8 input
\usepackage[T1]{fontenc}    % use 8-bit T1 fonts
\usepackage{hyperref}       % hyperlinks
\usepackage{url}            % simple URL typesetting
\usepackage{booktabs}       % professional-quality tables
\usepackage{amsfonts}       % blackboard math symbols
\usepackage{nicefrac}       % compact symbols for 1/2, etc.
\usepackage{microtype}      % microtypography
\usepackage{lipsum}
\usepackage{fancyhdr}       % header
\usepackage{graphicx}       % graphics
\graphicspath{{media/}}     % organize your images and other figures under media/ folder
\usepackage[square,numbers]{natbib}
\usepackage{longtable}

\defcitealias{source35}{CY}
\defcitealias{source20}{SM4T}
\defcitealias{source19}{ESCO}
\defcitealias{source36}{PYa}
\defcitealias{source22}{FACTc}
\defcitealias{source31}{iMOOC}
\defcitealias{source23}{SC}
\defcitealias{source25}{PROGa}
\defcitealias{source26}{PROGb}
\defcitealias{source27}{PROGc}
\defcitealias{source34}{PM}
\defcitealias{source29}{ICS}
\defcitealias{source32}{ML}
\defcitealias{source28}{HS}
\defcitealias{source24}{PYc}
\defcitealias{source37}{AP}
\defcitealias{new1}{FACTa}
\defcitealias{new2}{FACTb}
\defcitealias{new3}{PYb}

\title{A Systematic Literature Review of Computer Science MOOCs for K-12 education}

\author{
  L.M. van der Lubbe, S.P. van Borkulo, J.T. Jeuring \\
  Freudenthal Institute \\
  Universiteit Utrecht \\
  Utrecht, the Netherlands\\
  \texttt{l.m.vanderlubbe@uu.nl}}

\begin{document}
\maketitle

\begin{abstract}
Computer science (CS) is increasingly becoming part of the curricula of K-12 education in different countries. However, there are few K-12 CS teachers, and tools to offer K-12 CS education are often limited. Massive Open Online Courses (MOOCs) might help to temporarily address these challenges, and enable more schools to offer CS education. The goal of this systematic review is to give an overview of how CS MOOCs have been used in K-12 education. Nineteen papers from 2014 to May 2024 were included, describing thirteen different MOOCs. This review summarizes the research performed with these MOOCs and discusses directions for future research. Our findings show that most CS MOOCs target only part of the CS curriculum. When using a MOOC, a classroom teacher has an important role in supporting and managing students as they work in the MOOC. Research evaluating MOOCs is diverse, both in aims and in methods. 
In conclusion, MOOCs can play a valuable role in K-12 CS education, although additional teacher training to support students might be required. Moreover, additional learning material is needed to cover the full curriculum, as most MOOCs focus on programming and computational thinking. 
\end{abstract}


\section{Introduction}
The rapid pace of digital transformation is reshaping many aspects of our daily lives, influencing how we work, learn, and interact with the world around us \cite{source1}. In this evolving landscape, computational thinking (CT) has emerged as a crucial framework for navigating and harnessing these changes. CT refers to ``the thought processes involved in formulating a problem and expressing its solution(s) in such a way that a computer – human or machine – can effectively carry [it] out" \cite{source2}. In today's digital age, it is increasingly regarded as a fundamental skill, comparable to literacy and numeracy. Closely aligned with CT is computer science (CS), particularly programming, which equips individuals with the tools to understand and shape the world through technology. Recognizing the importance of these skills, many countries, including those in Europe, are incorporating CT and CS into their educational curricula \cite{source1, source7}. However, this integration faces significant challenges, such as a shortage of adequately trained educators and insufficient teaching tools and resources \cite{source1, shein2019cs}. 

This paper focuses on Massive Open Online Courses (MOOCs), which are an increasingly popular means to teach courses online to different audiences, although they mainly target adults. With the lack of CT and CS teachers and of tools and resources in secondary schools, maybe MOOCs can play a role in teaching CS, at least temporarily? We are aware of recent initiatives in this direction in the Netherlands (Co-Teach Informatica\footnote{\url{www.//co-teach.nl/}} \cite{van2023bridging}), Belgium (eTeacher\footnote{\url{www.eteacher.be}} \cite{eteacher}), Slovenia (Project Tomo\footnote{\url{www.projekt-tomo.si}} \cite{tomo}), Portugal (Ensico\footnote{\url{https://ensico.pt}}), and expect that in many other countries similar initiatives are set up. This review paper aims to give an answer to that question by giving an overview of the state-of-the-art of K-12 CS MOOCs.

\citet{source3} analysed trends in K-12 MOOCs. In their review, they studied 21 papers published between 2013 and March 2020. The majority of the papers found describe MOOCs for Science, Technology, Engineering, and Math (STEM) courses, including CS but also for example chemistry or physics. 
They found that the best way to use a MOOC in a classroom setting is a blended approach. 

In this paper, we focus solely on K-12 CS MOOCs and closely related topics such as CT and information literacy. Our analysis centers on course organisation (e.g.,~the duration of the course and hosting platform), their content, and the underlying pedagogical frameworks, with particular attention to the role of teachers. By providing this information, we aim to offer valuable insights for teachers, curriculum designers, and MOOC developers, helping them better understand how MOOCs can be integrated into secondary education. 

Additionally, we are interested in the research that has been conducted to design or evaluate the described MOOCs. For this reason, we restrict our review to academic publications about MOOCs. Researchers and MOOC designers can use the resulting overview to reuse methods for their own work, as well as to compare their findings with those of previous studies.

This paper is organised as follows. Section~\ref{sec:background} describes K-12 CS education and different relevant terms used to describe K-12 CS education. Moreover, this section describes MOOCs and their different variants, and their role in K-12 education.  Section~\ref{sec:methods} describes the methods we use for data collection and data analysis of our literature review. Section~\ref{sec:results} describes the results of the literature review, which are discussed in Section~\ref{sec:discussion}. Section~\ref{sec:conclusions} summarises the most important findings of this literature review, stressing their importance for future research. 

\section{Background}
\label{sec:background}
\subsection{K-12 Computer Science Education}
Computer science includes subjects such as programming, algorithms, data structures, information processing, and computer and network design \cite{source4}. Computer science is sometimes also referred to as computing or informatics \cite{source1}. Another term related to computing is computational thinking \cite{wing2006computational,source2}, which is a way of thinking that can be expressed in programming or coding, but also in unplugged situations. Computational thinking and CS are often closely related in education since learning CT skills is often achieved by learning CS \cite{source1}. The European Union has the ambition that all European citizens be educated in both digital literacy (how to use computers) and informatics (or computer science) \cite{source5}. While digital literacy is often integrated into educational curricula, teaching CS remains a challenge for many educators. Computing education at university level started many decades ago and has since expanded into K-12 education in many countries. The pace of this shift varies significantly by country \cite{source6, source7}. In some non-European countries, computer science has yet to enter the high school curriculum, while others have already incorporated it in primary education. 

Integrating computing education into the compulsory curriculum faces several challenges, including the lack of adequately trained teachers and the lack of tools and resources \cite{source1}. CS university students often do not consider a teaching career due to factors such as low salaries and the perceived nature of the job \cite{yeni2020or}. To reduce the teacher shortage, in-service teachers from other subjects can be trained and supported to teach CS \cite{flatland2022building, van2023bridging, quille2022building}.

When designing computer science education for K-12, a frequently used strategy is to rely on the existing pedagogical approach and curriculum of higher education and adapt this for younger students, for example by using a programming language that is perceived to be easier or omitting complex topics \cite{source6}. However, the pedagogical strategies that work for undergraduates often do not work for high-school students. \citet{source6} recommends that instead of adapting existing programs, it is better to start over and use age-appropriate pedagogical strategies. 

To promote diversity and equity in computer science, it is crucial to dispel stereotypes surrounding the field and its professionals. Additionally, \citet{source8} mention that policymakers should provide more support for K-12 CS curricula and prioritise inclusivity in the field. In that way, students will see more opportunities to pursue a CS career, contributing to an important field with a high job demand.

\subsection{MOOCs}
A Massive Open Online Course (MOOC) is \textquotedblleft an open-access online course (…) that allows for unlimited (massive) participation" \cite{source10}. MOOCs typically include interactive elements to engage learners. By 2021, MOOCs have reached 220 million learners worldwide (excluding China) \cite{source11}. In the same year, over 19,000 MOOCs were announced or launched by approximately 950 universities globally. The largest MOOC platforms in 2021 included Coursera, edX, FutureLearn, and Swayam. The most popular course categories in 2021 were Business (20.9\%) and technology (20.2\%), a trend that has remained consistent over time \cite{source11}. Furthermore, within the top 250 MOOCs of 2021, the largest category was “technology” comprising 42 courses, many of which focused on or were related to programming \cite{source12}. 

The defining characteristic of a MOOC is the possibility to serve a large number of participants. In contrast, Small Private Online Courses (SPOCs) are online courses that restrict access to a limited number of students \cite{source10}. There are different types of MOOCs, with different approaches to how they approach knowledge transfer. Extended MOOCs (xMOOCs) follow traditional lecture-based formats, where students have a passive role and the courses are structured over a fixed period \cite{source10}. These MOOCs are often used as supplementary material, extending traditional courses. When MOOCs include social media to encourage students to become active contributors to the course, they can be called connectivist MOOCs (cMOOCs) \cite{source10}. In cMOOCs, the emphasis is on personalized learning through a personalized learning environment, while xMOOCs are based on a more behaviourist pedagogical approach emphasising an individualistic learning approach \cite{source13}. A hybrid approach (hMOOC) combines formal and non-formal learning activities for example via an e-learning platform (xMOOC) with informal learning such as via a social network (cMOOC) \cite{source14}. 
Several taxonomies further classify MOOCs \cite{source13,source15}. MOOCs are also used in a so-called blended learning approach, in which traditional, face-to-face, learning activities are combined with online components from MOOCs \cite{source16}. \citet{source17} call this a bMOOC and propose a classification for how bMOOCs can be used. For example, they distinguish between a so-called integrated and supplementary model to distinguish MOOCs that are an integrated or mandatory part of education from supplementary and optional MOOCs. Within the integrated model, other models can be distinguished to further classify bMOOCs. 

\subsubsection{MOOCs for K-12 Education}
Although the majority of MOOCs target adults, MOOCs are becoming increasingly popular in secondary education \cite{source3}. However, it is not good practice to simply take an existing MOOC and use it with a K-12 audience. Instead, MOOCs for K-12 education need to be adapted to the way high school students learn \cite{source18}. It seems that the blended approach, embedding a MOOC in an existing school structure, is the most efficient way of using a MOOC in K-12 education \cite{source3}. Most MOOCs for K-12 education target STEM or CS topics and are designed to enhance or extend the current curriculum \cite{source3}.

\section{Materials and Methods}
\label{sec:methods}
This systematic literature review \cite{webster2002analyzing} focuses on MOOCs for K-12 CS education. More specifically, we focus on MOOCs for K-12 students, not for K-12 teacher training. Our main goal is to study academic publications of K-12 CS MOOCs to gain insight into the current state-of-the-art. While previously \citet{source3} conducted a broader review of all K-12 MOOCs, our analysis narrows in on papers related to CS education. From an exploration of the papers about CS education in their review, we identified recurring themes in the literature. We chose to describe the current state-of-the-art by focusing on two prominent topics: course organisation and content, and the research conducted.

The first topic, course organisation and content, helps to understand the types of MOOCs that are described in the literature. This includes both the practical side of teaching, such as course duration or hosting platforms, and the structure and delivery of the content to learners, including the didactical approach and the role of the teacher. Understanding how MOOCs are organized can help teachers to understand how MOOCs can be used, and can help (future) MOOC designers. Similarly, by understanding the content of MOOCs, teachers and designers get more understanding of which topics are already covered by MOOCs and for which topics it might be interesting to develop a new MOOC. As pointed out by \citet{source18}, the pedagogical approaches used in MOOCs for higher education cannot simply be transferred to MOOCs for K-12 education. Thus, it is important to study which pedagogical frameworks are specifically mentioned for K-12 education. Finally, in terms of course organisation and content, we want to verify whether the blended approach, identified as the most promising by \citet{source3}, is also the most suitable form for implementing MOOCs in K-12 CS education.

Thus, for the first part of the review, we investigate the following research questions (RQ):
\begin{enumerate}
    \item In what educational and technical settings are K-12 CS MOOCs implemented? 
    \item What content is covered in K-12 CS MOOCs? 
    \item What are the pedagogical frameworks used for these MOOCs?
    \item What role does the classroom teacher play in facilitating the MOOC?
\end{enumerate}

The second topic, which focuses on the research conducted, helps to understand the current research field of K-12 CS MOOCs. To get more insight into the scope of existing research, we examine the measures used and the characteristics of participants involved. This analysis can help to establish common research practices for others in this field. Moreover, our analysis of the conducted research can inform our future work section by identifying research gaps or shortcomings. For the second part of the review, we address the following additional research questions:
\begin{enumerate}
    \setcounter{enumi}{4}
    \item 
    Which research topics are studied in research towards K-12 MOOCs?
    \item What are the participant characteristics in studies performed with K-12 CS MOOCs?
\end{enumerate}

The following subsections describe the process of paper selection for this review and the method we use to analyse the selected papers.

\subsection{Paper selection}
For the paper selection, we used Scopus, ERIC, IEEE Xplore, Learning \& Technology Library, Web of Science and ACM DL using the following search keywords: ``MOOC'', ``MOOCs", ``Massive Open Online Course", ``Massive Open Online Courses", in conjunction with ``K12", ``K-12", ``secondary education", ``compulsory education", ``high school", ``middle school", and in conjunction with ``computer science", ``CS", ``computational thinking", ``CT", ``computing", ``informatics", ``digital literacy". We limited our results to the period 2014 - 2024 and to English articles or conference papers. Different search engines require slightly different queries, see Appendix \ref{app:query} for more details. The search was completed at the beginning of May 2024.

Table \ref{tab:queries} in Appendix \ref{app:query} shows the number of results for every search engine. Duplicates were removed before assessing the title and abstract of the papers to determine the inclusion of the papers in our study. Two researchers assessed the papers on the following inclusion criteria:
\begin{itemize}
    \item Published in peer-reviewed journal or conference paper;
    \item Papers are published after 2014;
    \item Papers are written in English;
    \item Papers focus on MOOCs used in K-12 education and address students;
    \item Papers discuss MOOCs for K-12 computer science education, including information literacy, and computational thinking;
    \item Papers focus on the students' experience or performance with the MOOC;
    \item MOOCs are specifically designed for K-12 education or have a program around it that is specifically target that age group.
\end{itemize}
Moreover, we excluded papers based on the following exclusion criteria:
\begin{itemize}
    \item Review papers;
    \item Papers that describe an online course that is not specified as MOOC;
    \item Papers that describe a case study with n=1;
    \item Papers that exclusively describe a design.
\end{itemize}
Disagreements among the researchers were resolved by reading the full papers.

Additionally, we checked the forward references (papers referencing the included papers) using Scopus and Google Scholar if an included paper was not available in the Scopus corpus. We also checked the references in the included papers for additional papers. In total, this resulted in a set of 19 papers that were used for further analysis in this review.

\subsection{Data analysis}
\label{sec:data-analysis}
After selecting the papers, we reviewed each included paper. Table \ref{tab:data-analysis} explains the categories we used for coding the included papers.

\newpage

\begin{table}[ht!]
\caption{Coding table for data analysis}
{\begin{tabular}{p{0.16\textwidth}p{0.22\textwidth}p{0.53\textwidth}} \toprule
 Topic & Research question & Coding \\ \midrule
 Overview of MOOCs (Table~\ref{tab:1}) &  & \textbf{Topic}: description
\newline \textbf{Implementation country}: country name \newline \textbf{Age range/school grade}: years old/grade(s)\\

  Overview of the research of the papers (Table \ref{tab:1b}) & & \textbf{Research aim}: categorical (Design description and evaluation; Compare user groups; Other + description) \newline \textbf{Research outcomes}: description
\\
\midrule
Course organization and content (Section \ref{sec:organization}) & 1: In what educational and technical settings are K-12 CS MOOCs implemented? (Section \ref{sec:q1}) & \textbf{Duration}: total or session duration \newline \textbf{Hosting platform}: platform name (Table \ref{tab:2b}) \newline \textbf{Online collaboration}: yes/no value for presence (Table \ref{tab:2b}) \\

& & \textbf{Learning materials}: yes/no value for presence of: video lectures, certificate of completion; other + description (Table \ref{tab:2c}) \\

& 2: What is the content of K-12 CS MOOCs? (Section \ref{sec:q2}) & \textbf{Programming language taught}: categorical (Scratch; Java; Python; Other) (Table \ref{tab:2d-alt}) \newline  \textbf{Sources of learning content}: existing curriculum (Table \ref{tab:3})  \newline \textbf{Expert collaboration}: yes/no value for presence (Table \ref{tab:3})\\

& 3: What are the pedagogical frameworks used for these MOOCs? (Section \ref{sec:q3}) & \textbf{Didactical foundations}: description (Table \ref{tab:3b})\\

& 4: What role does the classroom teacher play in facilitating the MOOC? (Section \ref{sec:q4}) & \textbf{Role of the classroom teacher}: description  \\

\midrule
MOOC research & 5: Which research topics are studied in research towards K-12 MOOCs? (Section \ref{sec:q5}) & \textbf{Measurements}: categorical (Student perceptions; Learning outcomes; Student behaviour; Other) (Table \ref{tab:4}) \newline  \textbf{Measurement instruments}: description (Table \ref{tab:4})  \\

& 6: What are the participant characteristics of the studies performed with the MOOCs? (Section \ref{sec:q6}) & \textbf{Participant numbers}:  Enrolled \& Included participant numbers (Table \ref{tab:5}) \newline \textbf{Gender ratio}: Male-female ratio (Table \ref{tab:5}) \\

  \bottomrule
\end{tabular}}
\label{tab:data-analysis}
\end{table}

\section{Results}
\label{sec:results}
\begin{figure}[ht!]
\centering
\includegraphics[width=0.55\textwidth]{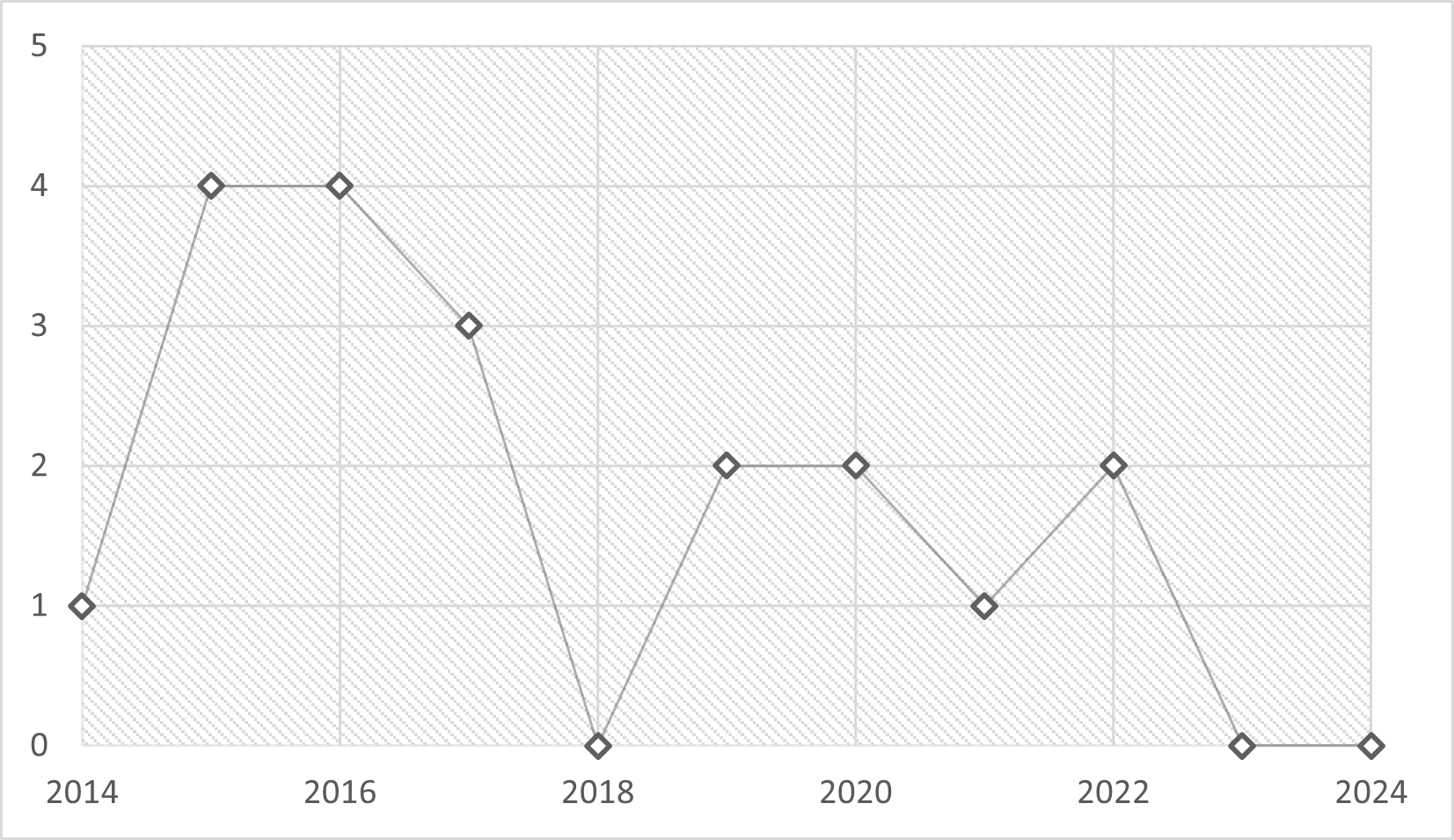}
\caption{Number of papers included per year} \label{fig:papers-years}
\end{figure}

We found 19 papers. Figure \ref{fig:papers-years} shows the number of papers per year.

The 19 papers describe a total of 13 MOOCs. To improve readability, we define short aliases to refer to every paper in the results section. These aliases and their corresponding reference can be found in Table~\ref{tab:1}, together with a brief description of the MOOC and the country in which the MOOC was implemented. 
Some MOOCs are offered in different variants, such as different language versions, updates in consecutive years, or variations based on the programming language taught. Different grades are specifically targeted, such as the 9th grade \cite{source29}, the 9th-10th grade \cite{source19}, the 11th-12th grade \cite{source32}, and the 3rd grade vocational education \cite{source26,source25,source27}. Note that for some MOOCs, the described age range is broader than just K-12. For example, \citet{source35,source34} mention that only a minority of their participants are within the K-12 age range. Although we restrict ourselves to papers that describe MOOCs for a K-12 target group, some MOOCs are not exclusively used by K-12 students. Two papers about MOOCs do not mention the targeted age range. They do mention ``high school" or ``K-12", but not the grades or ages \cite{source28, source37}. 

\begin{small}
  \begin{longtable}[c]{p{0.32\textwidth}p{0.45\textwidth}p{0.15\textwidth}
  }
  \caption{Overview of the MOOCs used in the included papers}
  \label{tab:1} \\
    \toprule
    \textbf{MOOC} & \textbf{Topic MOOC} & \textbf{Country} %&
    \\
    \midrule
    
    Scratch MOOC for Teens \cite{source20} [\citetalias{source20}] & Procedural thinking and problem-solving skills & Uruguay 
    \\ \midrule

    Code yourself \& A programar \cite{source35} [\citetalias{source35}] & Programming (English and Spanish), based on earlier experiences with SM4T MOOC. Fundamentals of programming, development of CT, basic programming practices in software engineering & UK and Uruguay	
    \\ \midrule
    
    ESCOMOOC14MA/IDAM \cite{source19} [\citetalias{source19}] & Mobile app development & Spain 
    \\ \midrule
    
    Foundations for Advancing Computational Thinking \cite{source22,new2,new1} [\citetalias{source22,new1,new2}] & Algorithmic thinking and programming, preparation for CT. 	& USA
    \\  \midrule
    
    iMOOC \cite{source31} [\citetalias{source31}] & Information literacy	& Switzerland	
    \\ \midrule
    
    Scratch MOOC by \citet{source23} [\citetalias{source23}] & Elementary programming concepts and software engineering concepts     & Netherlands	
    \\ \midrule
    
    PROG15/16 \cite{source26, source25, source27} [\citetalias{source26, source25, source27}] & National CS exam preparation& Greece 
    \\ \midrule
    
    MOOC by \citet{source34} [\citetalias{source34}] & Programming, designed for university level and used in secondary schools, can be used as entrance exam & Finland 
    \\ \midrule
    
    MOOC by \citet{source29} [\citetalias{source29}] & Learning program based on an existing ``Introduction to Computer Science" MOOC combined with offline activities used for science track classes	& Israel
    \\ \midrule
    
    Blended-learning Program on Machine Learning by \citet{source32} [\citetalias{source32}] & Blended program around Deep Learning Specialization MOOCs series\footnote{\url{https://www.coursera.org/specializations/deep-learning}}. Builds upon the earlier work of \citet{source29} & Israel	
    \\ \midrule
    
    MOOC by \citet{source28} [\citetalias{source28}] & MOOC teaching from the AP Computer Science A curriculum & USA
    \\ \midrule
    
    MOOC by \citet{source37} [\citetalias{source37}] & Preparation for the College Board’s Advanced Placement Computer Science A exam. Equivalent to an introductory CS course at college level	& USA
    \\ \midrule

    Python2014/2015 \& Java2017
    \cite{new3, source36, source24} [\citetalias{new3, source36, source24}] & Python (and Java) programming & Germany 
    \\
    
    \bottomrule
  \end{longtable}
%\end{table*}
\end{small}

Table \ref{tab:1b} gives an overview of the research that was reported on in the papers, summarising the main findings. Overall, the outcomes of the research on MOOCs for K-12 CS education paint a positive picture of the field. Most papers report findings that support the viability of using MOOCs for CS education, based on the student support, and learning or motivational outcomes. Some studies concentrate on more detailed research questions, such as the effect of online communication channels or differences between participant groups.

\begin{small}
  \begin{longtable}[c]{p{0.075\textwidth}p{0.35\textwidth}p{0.45\textwidth}}
  \caption{Overview of the research described in the included papers}
  \label{tab:1b} \\
    \toprule
    \textbf{Study} & \textbf{Research aim} & \textbf{Research outcomes}\\
    \midrule
       
    \citetalias{source20} \citeyear{source20} & Design description with evaluation & Positive evaluation of SM4T. It is a viable option to promote the development of procedural thinking and problem-solving skills using Scratch \\ \midrule

    \citetalias{source35}
    \citeyear{source35}& 	Design description with evaluation	& MOOC met or exceeded expectations of students and they would like to continue programming\\ \midrule
    
    \citetalias{source19} \citeyear{source19} & Study the effect of massiveness on the participation in social technologies & The ESCO MOOC has positive effects, as shown by an increase in the interactions of the participants \\ \midrule
    
    \citetalias{new1} \citeyear{new1} & Design description and evaluation & The blended MOOC worked as well as the face-to-face version. \\ \midrule

    \citetalias{new2} \citeyear{new2} & Design description and evalaution \newline Transfer of knowledge to other programming languages & Learning gains and transfer was achieved. Students also understand the computing discipline in a more mature way. \\ \midrule
    
    \citetalias{source22} \citeyear{source22} & Study the learning outcomes of the FACT MOOC and the factors that influence those outcomes & Overall knowledge gain. Prior computing experiences and math and English ability predicted learning outcomes. Extra-curricular experiences with technology also had an impact \\ \midrule

    \citetalias{source31} \citeyear{source31}  & Study the learning effect and student acceptance of a MOOC on information literacy & The student technology acceptance for a MOOC in K-12 education is low and is mainly driven by hedonic motivation, which negatively predicts learning gains \\ \midrule

    \citetalias{source23} \citeyear{source23} & Study whether students find programming concepts more difficult than software engineering concepts \newline Compare user groups: age \newline Predict successful completion & Both types of concepts score similarly \newline $>$12 y/o perform significantly better on particular topics \newline Student profile and behaviour in the first week predicts successful completion \\ \midrule

    \citetalias{source26} \citeyear{source26} & Design description and evaluation & Overall positive evaluation of the experience with the MOOC, learning expectations and general view on the use of MOOCs for education. Moreover, the paper summarizes lessons learned that can be used for the 2016 version of the course \\ \midrule

    \citetalias{source25} \citeyear{source25}  & Explore using a Facebook group as a complementary communication and collaboration platform & Facebook has the potential to play an important role in supporting collaborative learning communities for MOOCs \\ \midrule

    \citetalias{source27} \citeyear{source27} & Design description and evaluation & Positive results similar to other MOOC studies in higher education, such as comparable dropout rates and appreciation of the pedagogical approach of the MOOC and its learning benefits \\ \midrule

    \citetalias{source34} \citeyear{source34} & Compare user groups: age & No progress differences are found between younger and older participants \\ \midrule
    
    \citetalias{source29} \citeyear{source29} & Study whether participating students develop understanding of core computer science ideas and a more accurate perception of the nature of computer science & Participating students achieved results similar to undergraduate students and a more accurate perception of CS as a scientific discipline\\ \midrule
    
     \citetalias{source32} \citeyear{source32} & Design description and evaluation of MOOC and teacher training	& Initial results demonstrate achievement of learning goals and show that the MOOC might help teachers new to machine learning to teach this topic\\ \midrule

    \citetalias{source28} \citeyear{source28} & Compare user groups: face-to-face versus MOOC & A significant difference in student help-seeking abilities is found, but not in other self-regulated learning behaviours. Help-seeking is also the only strategy significantly related to MOOC students' grades on programming tasks\\ \midrule
    
    \citetalias{source37} \citeyear{source37} &	Determine the success of the MOOC \newline Compare user groups: with versus without coach & A MOOC is a viable option for high school students. Students with a coach performed better and were more active on the forum \\ \midrule

    \citetalias{source36} \citeyear{source36}  & Study the requirements for adapting a MOOC to support CS learning and teaching in secondary education \newline Compare user groups: age & MOOCs are suitable to use in secondary CS education. Social interaction and collaboration are important for successful participation \newline Age related differences in student behaviour and outcome \\ \midrule

    \citetalias{new3} \citeyear{new3} & Design description and evaluation & MOOC attracted a broad audience, and the MOOC was well received by the audience. Participants needed a lot of support \\ \midrule

    \citetalias{source24} \citeyear{source24} & Design description and evaluation & The study shows that MOOCs can be useful if CS is otherwise not offered at a school \\ 
    
    \bottomrule
  \end{longtable}

\end{small}

In the following subsections, we summarise the results for the different topics and research questions. If we could not find information on a topic in the paper(s) describing a MOOC, we do not mention the MOOC. Of course, that does not necessarily imply that something is missing in the MOOC or the corresponding research. 

\subsection{Course organisation and content}
\label{sec:organization}
We start by describing the way the MOOC was organised and the content of the MOOC. For this topic we have four research questions. For each of these questions, we will describe the aspects mentioned in Table \ref{tab:data-analysis}.

\subsubsection{RQ1: In what educational and technical settings are K-12 CS MOOCs implemented?}
\label{sec:q1}
As explained in Section~\ref{sec:methods}, we want to know how K-12 CS MOOCs are offered to learners by describing different aspects of the platform and the organisation of the MOOC. Therefore, in this section we first elaborate on the duration of the MOOCs. Next, we describe which hosting platforms are used, and whether there is an online communication component mentioned in the papers. Lastly, we focus on the type of learning materials mentioned in the papers (e.g.,~video lectures or quizzes), whether those are released all at once or in segments (if indicated), and whether users are provided with a certificate upon completion. 

\paragraph*{Duration.} 
Eight MOOCs (61\%) had durations of up to three months (12 weeks) [\citetalias{source35, source20,source19, source23, source25, source26}; \citetalias{source27}; \citetalias{source34, new1, new2, source22}; \citetalias{new3, source36}]. 
Among them, the shortest MOOC lasted only three weeks [\citetalias{source19}] and the longest extended to the full 12 weeks [\citetalias{source34}]. 

Some MOOCs were designed to be used for even a longer period. For instance, one of the MOOCs from \citetalias{source24}, lasted for half a school year, while earlier versions of the MOOC lasted for around a month. The \citetalias{source37} MOOC lasted a full year. The \citetalias{source28} MOOC consisted of four modules, each lasting six weeks. Additionally. the \citetalias{source32} MOOC spanned over two years, with the course running for eight months per year. Sometimes the duration of a MOOC was not expressed in weeks but as the total number of hours. For \citetalias{source31} for example, the total duration was around six hours.

Aside from the total duration, some MOOCs specify the hours spent per week. For example, \citetalias{source35} requires one hour per week from a student during six weeks. The \citetalias{source29} MOOC took 90 minutes per week and required an additional 30-45 minutes of homework; the number of weeks is not specified. For \citetalias{new1,new2, source22}, students met four times a week for periods of 55-minutes.

\paragraph*{Platform used and online collaboration.}
MOOCs are often offered on existing, large, platforms that host and distribute multiple MOOCs, such as Udemy, Coursera, and (Open) EdX. Some other MOOCs used more specific platforms. Table \ref{tab:2b} lists all platforms and indicates if online collaboration between MOOC students is facilitated. Either the hosting platform or additional (social media) platforms are used for this.

%\begin{table*}
%  \caption{Overview of the MOOCs used in the included papers}
%  \label{tab:1}
\begin{small}
  \begin{longtable}{p{0.2\textwidth}p{0.25\textwidth}p{0.18\textwidth}}
  \caption{Summary of used platforms and online collaboration}
  \label{tab:2b} \\
    \toprule
    \textbf{MOOC} & \textbf{Platform} & \textbf{Online collaboration} \\
    \midrule   
    
    \citetalias{source35} & Coursera &  \checkmark \\

    \citetalias{source20} & PlanCEIBAL's CREA platform & \checkmark \\
    
    \citetalias{source19} & Moodle & \checkmark\\
    
    \citetalias{new1, new2, source22}  & (Open) EdX & - \\
    
    \citetalias{source31} & (Open) EdX/ Swissmooc.ch & - \\

    \citetalias{source23} & (Open) EdX & \checkmark  \\
    
    \citetalias{source25, source26, source27} & Udemy and Facebook groups & \checkmark  \\
    
    \citetalias{source34}  & unknown & - \\

    \citetalias{source29} & unknown & \checkmark  \\
    
    \citetalias{source32} &  Coursera & - \\
    
    \citetalias{source28} & (Open) EdX & \checkmark  \\
    
    \citetalias{source37}  & unknown & \checkmark  \\  

    \citetalias{new3, source36, source24} & openHPI & \checkmark \\

    \bottomrule
  \end{longtable}
%\end{table*}
\end{small}

\paragraph*{Learning resources and certificates of completion.}
Most MOOCs (10; 77\%) mention that they offer video lectures, see Table \ref{tab:2c}. Often the video lectures feature a teacher explaining, but the lectures can also be in the form of for example screencasts [\citetalias{source35, source25, source26, source27}]. \citetalias{source31} uses storytelling, through videos, as well as levels and badges to add gamification to their MOOC. 

MOOCs offer assessments in different forms, such as assignments or quizzes. In some cases, these assessments are evaluated using auto-grading or peer assessment. Some MOOCs offer automatic grading of code through special programming environments [\citetalias{source34, source24, new3}].

Six MOOCs hand out a certificate of completion or participation after finishing the MOOC, see Table \ref{tab:2c}. To obtain such a certificate, MOOCs often require a certain percentage of completed assessments or earned points. 

%\begin{table*}
%  \caption{Overview of the MOOCs used in the included papers}
%  \label{tab:1}
\begin{small}
  \begin{longtable}{p{0.2\textwidth}p{0.12\textwidth}%p{0.10\textwidth}
  p{0.35\textwidth}%p{0.11\textwidth}
  p{0.18\textwidth}}
  \caption{Summary of learning resources}
  \label{tab:2c} \\
    \toprule
    \textbf{MOOC} & \textbf{Video lectures} & %\textbf{Assign-ments} & 
    \textbf{Other}  & % \textbf{Segmental release} &
    \textbf{Completion certificate}\\
    \midrule
    
    \citetalias{source35} & \checkmark & -  & \checkmark  \\

    \citetalias{source20} & \checkmark & -  & \checkmark \\
    
    \citetalias{source19} & \checkmark & -  & - \\
    
    %\citetalias{source33}  & - & \checkmark & - & - & - \\

    \citetalias{new1, new2, source22} & \checkmark & -   & - \\
    
    \citetalias{source31} & \checkmark & Gamification  & - \\

    \citetalias{source23}  & \checkmark & - & - \\
    
    \citetalias{source25, source26, source27}  & \checkmark & - & -\\
    
    \citetalias{source34}  & -  & Programming environment with auto-grading & -  \\

    \citetalias{source29} & - & - & - \\
    
    \citetalias{source32} & - & -  & \checkmark \\
    
    \citetalias{source28} & \checkmark  & - \\
    
   \citetalias{source37} & \checkmark & -  & - \\  

   \citetalias{new3, source36, source24} & \checkmark & Programming environment with auto-grading & \checkmark \\

    \bottomrule
  \end{longtable}
%\end{table*}
\end{small}

\subsubsection{RQ2: What is the content of K-12 CS MOOCs?}
\label{sec:q2}
This paragraph focuses on the second research question, which explores the nature of the content of the MOOC. First, we enumerate the programming languages taught in the reviewed MOOCs. Next, we elaborate on the sources that are used for the content of the MOOC. In some cases existing curricula or courses are used, others mention that the materials are made in collaboration with experts. Together, this information helps to get an overview of what the MOOCs are about.

\paragraph*{Programming language.}
Most papers (14) specify the programming language taught in their MOOC(s). Both text-based languages (Python and Java) and block-based language Scratch are used in the MOOCs. Text-based languages are used slightly more (6 MOOCs) compared to block-based (4 MOOCs). See Table \ref{tab:2d-alt}\footnote{The total of this table is 14 instead of 13. One MOOC teaches two different programming languages.} for the MOOCs per programming language.

%\begin{table*}
%  \caption{Overview of the MOOCs used in the included papers}
%  \label{tab:1}
\begin{small}
  \begin{longtable}{p{0.2\textwidth}p{0.18\textwidth}p{0.5\textwidth}}
  \caption{Programming languages taught or used in MOOCs}
  \label{tab:2d-alt} \\
    \toprule
    \textbf{Programming language} & \textbf{Number of MOOCs} & \textbf{MOOCs} \\
    \midrule
    \textbf{Scratch} & 4 & \citetalias{source35, source20, new1}, \citetalias{new2}, \citetalias{source22, source23} \\

    \textbf{Python} & 3 & \citetalias{source29, source32}, \citetalias{source36, source24, new3} \\

    \textbf{Java} & 3 & \citetalias{source34, source24, source37}\\

    \textbf{Not mentioned} & 4 & \citetalias{source19, source31},\citetalias{source26},\citetalias{source25},  \citetalias{source27, source28}  \\
    \bottomrule
  \end{longtable}
%\end{table*}
\end{small}

\paragraph*{Sources of learning content.} 
Six MOOCs are based on an existing curriculum or existing learning materials. Table~\ref{tab:3} lists the sources that these MOOC are based on. The last column of this table shows that two MOOCs mention that the learning materials are created in collaboration with (university) teachers, scientists, and students [\citetalias{source35, source20}] instead of mentioning a specific existing curriculum. The papers on the other MOOCs do not specify the sources of the learning content. 

\begin{small}
  \begin{longtable}{p{0.2\textwidth}p{0.55\textwidth}p{0.2\textwidth}}
  \caption{Sources of learning materials}
  \label{tab:3} \\
    \toprule
    \textbf{MOOC} & \textbf{Based on existing material} & \textbf{Created with experts} \\
    \midrule
    \citetalias{source35} & - & \checkmark \\

    \citetalias{source20} & - & \checkmark  \\
    
    \citetalias{source19} & - & -  \\
    
    \citetalias{source22, new1, new2} & Exploring Computer Science high school curriculum \cite{source45} & -\\
    
    \citetalias{source31} & - & - \\

    \citetalias{source23} & - & - \\

    \citetalias{source25, source26, source27} & Greek curriculum and analysis of past national exams & - \\
    
    \citetalias{source34} & CS1 university course at the University of Helsinki & - \\
    
    \citetalias{source29} & Academic MOOC introduction to computer science (First Steps in Computer Science and Programming in Python) developed by a university & -\\
    
    \citetalias{source32} & Content from popular MOOCs of Andrew Ng & - \\
    
    \citetalias{source28}& - & - \\
    
    \citetalias{source37} & Advanced Placement Computer Science; equivalent to a semester-long college introductory course & - \\ 

    \citetalias{new3,source36, source24} & - & - \\

    \bottomrule
  \end{longtable}
\end{small}

\subsubsection{RQ3. What are the pedagogical frameworks used for these MOOCs?}
\label{sec:q3}
To answer the research question about pedagogical frameworks used for developing MOOCs, we looked at theories, principles, and frameworks cited in the papers. Learning by doing is mentioned in one paper; another paper uses a citation for this. We only include a citation when it is mentioned in the paper. Table \ref{tab:3b} summarizes the pedagogical frameworks we found. 

\begin{small}
  \begin{longtable}{p{0.2\textwidth}p{0.4\textwidth}}
  \caption{Summary of didactical foundations}
  \label{tab:3b} \\
    \toprule
    \textbf{MOOC} & \textbf{Didactical foundations}\\
    \midrule
         
    \citetalias{source35} & Revised version of Bloom’s Taxonomy \citep{source47} \newline Community of Inquiry framework \citep{source48} \\

    \citetalias{source20} & -  \\
    
    \citetalias{source19} & MOOC principles by \cite{source40} \\
    
    %\citetalias{source33} & - \\

    \citetalias{source22,new1,new2} & Learning by doing \citep{barron2008can} \newline Cognitive apprenticeship \citep{source46} \newline Scaffolding \citep{pea2004social} \\
    
    \citetalias{source31} &  MOOC is designed according to \citep{source50} \\

    \citetalias{source23} & Bloom’s taxonomy \citep{source49}  \\

    \citetalias{source25, source26, source27} & - \\
    
    \citetalias{source34} &  Extreme Apprenticeship \citep{source42, source41} \\
    
    \citetalias{source29}  & Community of Inquiry framework \citep{source48} \\
    
    \citetalias{source32}  & Community of Inquiry framework \citep{source48}  \\
    
    \citetalias{source28} & - \\
    
    \citetalias{source37} & - \\  

    \citetalias{new3, source36, source24} &  Learning by doing \newline Constructivist approach \newline Bloom’s taxonomy \citep{source49} \\

    \bottomrule
  \end{longtable}

\end{small}

The design of the \citetalias{source19} MOOC uses principles formulated by \citet{source40}. These principles are: (1) competence-based design approach, (2) learner empowerment, (3) learning plan and clear orientations, (4) collaborative learning, (5) social networking, (6) peer assistance, (7) quality criteria for knowledge creation and generation, (8) interest groups, (9) assessment and peer feedback, and (10) media-technology-enhanced learning. 

The \citetalias{source34} MOOC incorporates the pedagogical method of Extreme Apprenticeship \cite{source42, source41}, which is particularly suitable for programming education. It emphasises learning by doing and continuous feedback and practice until a pupil masters the content. In the \citetalias{source34} MOOC, students work in a real-world programming environment from the start. During the course, students develop programs for numerous problems.

Two other MOOCs (\citetalias{new1, new2, source22, source24}) explicitly mention that learning by doing is an important principle incorporated in their MOOCs, without mentioning the Extreme Apprenticeship method. The FACT MOOC refers to the work of \citet{barron2008can} to support this choice. In addition, the FACT MOOC also shows how computer science was applied in different fields together with scaffolding \cite{pea2004social} and cognitive apprenticeship \cite{source46} by showing real-world programming examples. 

\citetalias{source24} mentions a social constructivist approach, following Vygotsky’s theory of proximal development \cite{vygotsky1978development}.

\citetalias{source31} uses the criteria for high-quality MOOCs of \citet{source50} for its design and testing. \citet{source50} introduces a rating instrument with 20 items in 15 categories in three sections to assess the instructional quality of MOOCs. The sections are (A) structuredness and clarity, (B) first principles of instruction, and (C) additional principles of instruction. Section A includes questions about the quality of the descriptions of learning goals, audience, requirements/effort, course contents, and course structure. Section B assesses the following principles: problem-centeredness, activation, demonstration, application, and integration. Finally, section C assesses the feedback, authenticity of resources, differentiation in the course, cooperation/collaboration in the course, and learner/activity orientation of the course.  

Three MOOCs (\citetalias{source35, source29, source32}) use the Community of Inquiry framework \cite{source48} to explain the design of their MOOC. In this framework, the Community of Inquiry consists of teachers and students. The model of the Community of Inquiry assumes that learning occurs in this community through the interaction of social presence, teacher presence and cognitive presence. 

Bloom’s Taxonomy classifies educational learning objectives on different levels \cite{source49}. \cite{source47} revise the taxonomy into a two-dimensional framework. Three MOOCs (\citetalias{source35, source23, source24, new3}) mention the use of Bloom’s Taxonomy or the revised version when designing the course and its assessments.

\subsubsection{RQ4: What role does the classroom teacher play in facilitating the MOOC?}
\label{sec:q4}

Although MOOCs are typically designed for individual learning, nine K-12 MOOCs are implemented in schools in a blended learning setting, where digital components are combined with classroom activities [\citetalias{source20, source19, source22, source31, source32, source29, source37, new1, new2}; \citetalias{source24,source36}]. Not all MOOCs specify the role of the teacher within this blended learning setting. 

\citetalias{source31} is a blended MOOC, in which an online MOOC is combined with in-class instruction and interaction. Teachers receive technical support and pedagogical advice upon request.

High schools can choose to purchase an additional coaching package for the \citetalias{source37} MOOC. The school then appoints a coach for a group of students using the MOOC, and the coach receives additional materials in the purchased package. 

The blended learning program of \citetalias{source29} combines online and offline activities. Teachers receive a Teacher Toolkit to help them to support social aspects and stimulate self-learning and collaboration. These are also important 21st-century skills that, according to the researchers of \citetalias{source29}, usually receive less attention in CS classes. In a similar MOOC (\citetalias{source32}), the role of the teacher is again to motivate students and to pay attention to the mental processes of the students and social processes of the class. Moreover, they introduce learning activities and discuss those activities with the classroom afterwards. Teachers are not the first help-seeking point for students. Instead, students have to turn to their peers for help. The interaction between students, the MOOC, and the teachers is based on the Community of Inquiry framework \cite{source48}. Teachers working with this MOOC are offered training and materials to help to develop a new mindset in their classroom and assist their students.

Some school teachers that are associated with MOOCs lack relevant content knowledge. Their role is mainly in classroom management [\citetalias{source20, source22, new1, new2}].

\subsection{MOOC research}
\label{sec:research}
In this section we explore the research that is described in the papers included in this review. The two research questions that are discussed in this section are about the measures that are used and the participant characteristics. A general description of the research and the outcomes can be found in Table \ref{tab:1b}.

\subsubsection{RQ5: Which research topics are studied in research towards K-12 MOOCs?}
\label{sec:q5}
This section summarises the results on the MOOC research measurements, see Table \ref{tab:4}. We divide the papers in four categories of measures, namely: student perceptions, learning outcomes, student behaviour, and other measures. Student behaviour includes interaction with the MOOC, social interactions and learning behaviour (such as asking questions). 

For the four categories we describe which measurement instruments were used. If we use the term ``MOOC data" we mean all data collected in and produced by the MOOC, including completion rates, scores, and interactions. If we mention ``student feedback" we mean any form of feedback given by students, which can be for example feedback from a survey or interviews. The same holds for ``teacher feedback". Only if specific questionnaires are mentioned, we mention those.

In the following paragraphs, we highlight interesting findings for every type of measure that we identify.

\begin{small}
  \begin{longtable}{p{0.1\textwidth}p{0.17\textwidth}p{0.64\textwidth}}
  \caption{MOOC research measurements and instruments summary}
  \label{tab:4} \\
    \toprule
    \textbf{Study} & \textbf{Measurements} & \textbf{Measurement instruments}\\
    \midrule
    
    \citetalias{source35} \citeyear{source35}  & Student perceptions \newline Student behaviour & MOOC data \newline Student feedback \newline Observations \\

    \citetalias{source20} \citeyear{source20}  & Student perceptions \newline Other & MOOC data \newline Student feedback \newline Teacher feedback \\
    
    \citetalias{source19} \citeyear{source19}  & Student behaviour & MOOC data\\

    \citetalias{new1} \citeyear{new1} & Learning outcomes & Computational Knowledge test (based on \citet{source54}, \citet{meerbaum2010learning} and \citet{source55}) \newline Preparation for Future Learning Test (inspired by \citet{schwartz2004inventing}) \newline Prior Experience Survey (adapted from \citet{barron2004learning}) \newline CS Perceptions Survey (based on \citet{source54}) \newline Online Learning Experience Survey (inspired by \citet{barron2004learning}) \newline FACT Experience Survey \newline Observations \\

    \citetalias{new2} \citeyear{new2} & Learning outcomes & Computational knowledge test (based on \citet{source54}, \citet{meerbaum2010learning} and \cite{source55}) \newline Preparation for Future Learning Test (using AP CS questions, inspired by \citet{schwartz2004inventing}) \newline Prior Experience Survey (adapted from \citet{barron2004learning}) \newline CS Perceptions Survey (based on \citet{source54}) \\
    
    \citetalias{source22}   \citeyear{source22} & Learning outcomes & Prior experience questionnaire \newline CS Interest \& Attitudes questionnaire (inspired by \citet{source54}) \newline A measure of Computational Thinking (based on \citet{source54} and \citet{source55}) \newline Student feedback \\

    \citetalias{source31}  \citeyear{source31}  & Student perceptions \newline Learning outcomes & Student feedback including the unified theory of acceptance and use of  technology (UTAUT(2), \cite{source56, source57}) adapted to the context \newline Information literacy performance test \cite{source58} \& MOOC data \\

    \citetalias{source23}  \citeyear{source23}  & Learning outcomes \newline Student behaviour & MOOC data \\

    \citetalias{source26}   \citeyear{source26}  & Student perceptions \newline Student behaviour & Student feedback \newline MOOC data \\

    \citetalias{source25}   \citeyear{source25}  & Student behaviour & MOOC data \\

    \citetalias{source27}  \citeyear{source27}  & Student perceptions \newline Student behaviour & Student feedback \newline MOOC data \\
    
    \citetalias{source34}   \citeyear{source34}  & Student perceptions \newline Student behaviour & MOOC data \newline Student feedback \\

    \citetalias{source29} \citeyear{source29}  & Learning outcomes \newline Other & Summative exam \newline Questionnaire on perceptions of the discipline of computing\\
    
    \citetalias{source32} \citeyear{source32}  & Student perceptions \newline Learning outcomes \newline Other & Student feedback \newline Teacher feedback \newline MOOC data\\
    
    \citetalias{source28}  \citeyear{source28}  & Learning outcomes \newline Other & Subscales from the Motivated Strategies for Learning Questionnaire \cite{source53} \newline MOOC data \\
    
    \citetalias{source37} \citeyear{source37}  & Learning outcomes \newline Student behaviour & Advanced Placement exam \newline MOOC data \\  

    \citetalias{source36} \citeyear{source36}  & Learning outcomes \newline Student behaviour \newline Other & Student feedback \newline Teacher feedback \newline MOOC data \\

    \citetalias{new3} \citeyear{new3} & Student perceptions \newline Student behaviours & Student feedback \newline MOOC data \\

    \citetalias{source24} \citeyear{source24}  & Student perceptions \newline Student behaviour \newline Other & MOOC data \newline Teacher feedback \newline Student feedback \\

    \bottomrule
  \end{longtable}
\end{small}

\paragraph*{Student perceptions.}
Most researchers (nine papers) ask students to fill out a survey to gain insights into student perceptions [\citetalias{source35, source20, source31, source26, source34, source27, source24, new3, source32}]. Most papers do not mention whether or not these surveys are based on other surveys, nor what the exact content of the surveys is. \citetalias{source31} uses the Unified Theory of Acceptance and Use of Technology (UTAUT(2), \cite{source56, source57}) adapted to the context of iMOOC to evaluate student perceptions. The authors try to determine predictors of technology acceptance.

In some cases surveys have a particular focus, such as the experienced assignment difficulty and educational value [\citetalias{source34}], or the expectations of the students [\citetalias{source26, source27}]. The research on \citetalias{source24} pays special attention to the team activity included in the MOOC. The evaluation of \citetalias{source27} specifically looks at the difference between students that face learning disabilities and those who do not. Some surveys are conducted both before and after the MOOC to see if student perceptions change. One study mentions interviews conducted with students and teachers [\citetalias{source20}], and another mentions student feedback but does not mention how that is collected [\citetalias{source32}]. 

Eight of the nine studies involving student perceptions report positive findings on aspects of these perceptions [\citetalias{source35, source20, source26, source32, source27, source24, source34, new3}]. The study of \citetalias{source34} finds that the experienced educational value of the exercises of the MOOC increase during the MOOC, i.e.~exercises at the end are rated higher compared to those in the beginning. Moreover, they find a significant difference in the experienced educational value between the two age groups in their study (15-19 and $>$20 years old), where the younger group considers this lower. Not all results are positive: the student acceptance of \citetalias{source31} is low. 

Surprisingly, one paper mentions collecting course experience via a survey but does not report on the results [\citetalias{source22}]. Another paper mentions surveys but does not describe the content or purpose of the surveys [\citetalias{source36}].

\paragraph*{Learning outcomes.}
Five studies measure learning outcomes by using scores derived from the collected MOOC data [\citetalias{source36, source23, source32,source28, source37}]. Three studies using MOOC data also explore which factors influence or predict learning outcomes [\citetalias{source31, source23, source37}]. Other MOOCs use specific measures such as exams or questionnaires to measure the learning outcomes of the MOOC, see all details in Table \ref{tab:4}.

\paragraph*{Student behaviour.}
Student behaviour is inferred from collected MOOC data in 11 papers [\citetalias{source35, source19, source23, source25, source26, source27, source37}; \citetalias{source34}; \citetalias{source36,new3,source24}]. Ten papers report the completion rate (percentage of students that complete the MOOC) [\citetalias{source35, source19, source31, source23, source26, source27}; \citetalias{source34}; \citetalias{new3,source36,source24}]. Two studies with a different research aim report completion rate on the side [\citetalias{source20, source32}]. In one study completion is not considered to be binary, but calculated as the percentage of the course completed by a participant [\citetalias{source37}]. Other MOOC data studies report are, amongst others, forum interactions [\citetalias{source19}] and social media data [\citetalias{source25}]. The research of the \citetalias{source23} MOOC aims to predict whether or not a student will complete a MOOC by looking at their profile and activities in the first week.

Completion rates are often compared to the existing literature on completion/drop-out rates in MOOCs, in which completion rates below 10\% are common \cite{source59}. Seven of the reviewed studies report higher completion rates compared with the existing literature [\citetalias{source20, source19, source31, source34, source32}; \citetalias{source36, source24}], while five others report similar rates [\citetalias{source35, source23, source26, source27}[ \citetalias{new3}]].

For social interactions, via forum(s) or social media (groups), the results are mixed. While four studies report those features were successfully used  [\citetalias{source19, source23, source25, new3}], two other papers report mixed results [\citetalias{source36, source24}]. A positive example is a study showing that using a forum results in a significantly higher post-test performance score [\citetalias{source37}]. In the \citetalias{source36} MOOC the forum is more often used when students receive their instruction at home compared to those who attend classroom instructions, a mixed result. Similarly, the study of \citetalias{source24} finds that when students work in groups and/or receive support from their teacher they are less active on the forum. In the study of \citetalias{new3}, the forum interactions are studied to consider the possibilities for scaling the support in the MOOC. Using the demographics from their MOOC and a similar MOOC, they study the differences between those MOOCs in terms of support needed by the participants. One of the findings is that their MOOC serves more novice participants compared to the other MOOC, which might explain the higher number of forum posts. 

In experiments with the \citetalias{source35} MOOC researchers observed students. Technical difficulties that students face are discussed as lessons learned that need to be taken into account for future implementations of the MOOCs. Moreover, the way that students interact with the forum showed a vibrant community of students. However, there is a need for instructions to make them follow threads on the forum and reduce redundancy.

\paragraph*{Other measures.}
Some studies focus on aspects that cannot be captured by one of the other categories. For example, some interview, or distribute surveys among, teachers or school administrators about their experiences and opinions about MOOCs, and some discuss ways to include MOOCs in K-12 education [\citetalias{source20, source32}; \citetalias{source36, source24}]. 

Subscales from the Motivated Strategies for Learning Questionnaire questionnaire \cite{source53} are used to study the self-regulated learning of the \citetalias{source28} MOOC participants.

The work on the \citetalias{source29} MOOC studies, using questionnaires, the perceptions of participants of the discipline of computing. 

\subsubsection{RQ6: What are the participant characteristics of the studies performed with the MOOCs?}
\label{sec:q6}
In this section, we present the participant numbers and gender to give an overview of the scope of the studies that are performed in the included papers (see also Table \ref{tab:5}).

With respect to the number of participants, we distinguish between the enrolled participants and the included participants. Enrolled participants are all participants that are registered in the MOOC. The included number of participants is the number of students that start or complete the MOOC and are used for analysis, which is often lower than the enrolled number of participants. With respect to the gender ratio, we only differentiate between male and female genders, as most papers report this. If the percentages mentioned do not add up to 100\%, the other participants either did not specify their gender or did not identify with male or female.

\paragraph*{Number of participants.}
The first two columns of Table \ref{tab:5} list the number of enrolled and included participants. In most cases, the number of included participants is smaller than the number of enrolled participants. Mostly, in studies less than half of the included number of participants are considered as enrolled participants. Exceptions are \citetalias{source20}, where 58.1\% of the enrolled participants start the course. However, only 18.6\% of the enrolled participants finished the course. Another exception is the \citetalias{source23} MOOC,  where 69.8\% of the enrolled participants are considered active users of the MOOC, however also this MOOC reports smaller numbers of users with profile information and even smaller numbers of users that complete the course. The \citetalias{source25, source26} MOOCs report that slightly over half of the participants (51.9\%) are considered active participants. For the Python 2017 MOOC from \citetalias{source24}, 58.1\% of the enrolled participants receive the Confirmation of Participation. Again, this MOOC shows lower numbers of participants with the Record of Achievement and with a completed post-course survey. Moreover, the other MOOCs from \citetalias{source24} also consider less than 50\% of their enrolled participants for analysis. The MOOC with the lowest dropout between enrolled and included is the \citetalias{source31}: 85.2\% of the enrolled participants filled in the pre- and post-test. However, even for this MOOC only 16.2\% of the enrolled participants completed questionnaires for the study.

\paragraph*{Gender ratio.}
Computer science is often a male-dominated field. To determine if this is also the case for the MOOCs included in this review, we study the gender ratios that are reported in the reviewed papers, which can be found in the last column of Table \ref{tab:5}. We only report on male-female ratio, as the majority of the studies only use this binary gender distinction.

Just four papers report almost equally large gender groups for their MOOCs (ratio between 40-60 and 50-50) [\citetalias{source35,source20,source31, source26}]. Ten papers (covering six MOOCs) report more unequal gender ratios, in which female students are a minority [\citetalias{source35, source22, new1, new2,source23, source34,source28}; \citetalias{source36, source24, new3}]. The reported percentages of female participants varies between 5.2\% and 37\%.

\begin{small}
  \begin{longtable}{p{0.08\textwidth}p{0.2\textwidth}p{0.28\textwidth}p{0.33\textwidth}}
  \caption{Overview of participant characteristics of the MOOC research}
  \label{tab:5} \\
    \toprule
    \textbf{Study} & \textbf{Enrolled participants} & \textbf{Included participants} & \textbf{Gender ratio}\\
    \midrule
    
    \citetalias{source35} \citeyear{source35}  & 84786 (total of 2 MOOCs) & 2483 completed final survey (total of 2 MOOCs) & "Code yourself": 54\% male - 44\% female  \newline "A Programar": 65\% male - 34\% female  \\ \midrule

    \citetalias{source20}  \citeyear{source20} & 1293 & 752 started the course \newline 241 (32.05\%) finished & 49\% male - 51\% female (enrolled participants) \\ \midrule
    
    \citetalias{source19}  \citeyear{source19} & 309 (247 MOOC, \newline 62 control) & - & - \\ \midrule

    \citetalias{new1} \citeyear{new1}; \citetalias{new2} \citeyear{new2}; \citetalias{source22}  \citeyear{source22} &  54 (group 1 = 26, \newline group 2 = 28) & -  & Group 1: 80.8\% male - 19.2\% female \newline Group 2: 71.4\% male - 28.6\% female\\ \midrule

    \citetalias{source31}  \citeyear{source31}  & 1032 & 880 pre- and post-tests \newline 167 questionnaires & 51\% female \\ \midrule

    \citetalias{source23}   \citeyear{source23} & 3179 & 2220 active students \newline 1243 with profile information \newline 181 completed the course & 31,66\% with gender on profile is female  \\ \midrule

    \citetalias{source26}  \citeyear{source26} & PROG15: 291 & PROG15: 151 active students & 59\% male - 41\% female; based on the before-PROG15 questionnaire\\  \midrule

    \citetalias{source25}   \citeyear{source25}& PROG15: 291 \newline PROG16: 265 & PROG15: 151 active students \\ \midrule

    \citetalias{source27} \citeyear{source27} & PROG15: 291 \newline PROG16: 433 & - & - \\ \midrule
    
    \citetalias{source34} \citeyear{source34} & 2109 & 358 answered web survey & 15-19 years old: 91.7\% male - 5.2\% female \newline $\geq$ 20 years old: 85.3\% male – 12.8\% female\\ \midrule
    
    \citetalias{source29} \citeyear{source29} & 55 & - & - \\ \midrule
    
    \citetalias{source32}  \citeyear{source32} & 24 \newline 273 (extension of the study, the results of which are not discussed) & - & -\\ \midrule
    
    \citetalias{source28}  \citeyear{source28} & 72 (31 face-to-face, 41 MOOC) & - & Face-to-face: 96.8\% male - 0\% female \newline MOOC: 36.6\% male - 22\% female\\ \midrule
    
    \citetalias{source37}  \citeyear{source37} & 5692 & 1613 students took the AP exam & - \\  \midrule

    \citetalias{source36}   \citeyear{source36} & $>$ 20000 (total of 2 years) & +/- 6000 surveys (total of 2 years) & $<$20 years old students: 25\% female  \newline $\geq$20 years old students: $<$20\% female\\ \midrule

    \citetalias{new3} \citeyear{new3} & +/- 7400 & 25.6\% profile data & 77.2\% male - 22.8\% female (based on profile data) \\ \midrule

    \citetalias{source24} \citeyear{source24} & Python 2017: +/- 1000 \newline Python 2018: 312 \newline Java 2018: 500 & Python 2017: 431 (Record of Achievement (RoA)), 581 (Confirmation of Participation (CoP)), 209 post-course survey \newline Python 2018: 87 RoA, 131 CoP, 146 pre-course surveys (no post-survey conducted) \newline Java 2018: 83 RoA, 139 CoP, 213 pre-course survey & Python 2017/2018: +/- 70\% male - 30\% female \\ 

    \bottomrule
  \end{longtable}
\end{small}

\section{Discussion}
\label{sec:discussion}
This review gives an overview of MOOCs for K-12 CS education and their key characteristics and use. This is a relatively small field, with only 19 relevant papers published in the last ten years. There are quite some differences between these MOOCs: they are based on different curricula, have different lengths, target audiences, age ranges and so on. We find some aspects that are shared among the MOOCs as well. Often, MOOCs focus on specific aspects of the curriculum, mostly programming or computational thinking, while languages for web development or database management (such as HTML, CSS, and SQL) are not mentioned. This narrow focus is reflected in the typically short duration of these MOOCs, mostly lasting no more than three months.

The background section describes the different types of MOOCs (xMOOC, cMOOC, hMOOC), using their pedagogical approach for classification. Furthermore, MOOCs can be used in blended learning settings, sometimes also called bMOOCs. Only two included papers use the terms x-, c-, or hMOOC to describe a MOOC \cite{source26, new3}. In addition, the term hybrid is used ambiguously. For example, \citet{source37} call their MOOC hybrid because it includes instructor intervention in the form of coaching and online forum instructor responses. \citet{source26} calls their MOOC hybrid since it combines the learning resources on one platform with asynchronous collaboration platforms. 

The included papers often focus on the content of a MOOC, and sometimes also on its translation to learning materials. They rarely discuss technical aspects and design choices, and the MOOCs often use existing platforms. This implies that most MOOC designers mainly focus on content while using existing infrastructures. We found one paper that focused on the use of an integrated Python programming environment \cite{new3}. Innovations such as personalisation of learning paths are not explored in the research included in this review. 

Taking the technological aspects of MOOCs into account is important. The research around the “Code yourself/A programar”-MOOCs finds that students have fewer competencies in basic computer skills (such as downloading and saving files) than expected \cite{source35}. Assuming a certain competence might thus be a pitfall, and it might be advisable to include instructions on important basic skills. \citet{new3} studies the performance of the load on the cloud that hosts their programming environment, to test the technical performance of their MOOC for scalability. Another technical aspect that should be considered is  accessibility. One of the studies mentions that the developers of the MOOC found that the MOOC was difficult to use for colour-blind users \cite{source35}. Accessibility can also be related to the form in which the learning materials are offered. The research of \citet{source22} shows that the level of math and English are related to learning outcomes. It would be interesting to study if different types of learning materials (for example text-based versus video-based) could mitigate the effect of the level of English on the learning outcomes. Such research might be useful for increasing the inclusivity of (CS) MOOCs. 

However, besides the examples mentioned in the previous paragraph, the evaluations show little focus on the technical aspects of MOOCs. Nine papers discuss student perceptions, but none of these focused on perceptions of  technical aspects. One paper discusses  student acceptance of the MOOC based on the UTAUT, which evaluates a MOOC as a whole, and only says something about the technology at an abstract level \cite{source31}.

The content and design of MOOCs have been described in different ways in the reviewed papers. Some papers only briefly mention the topic of the course and mainly focus on the study they perform with that MOOC. Other papers explain the rationale behind the design of their content and learning materials in more detail. Giving insight in the design of content is valuable for others working on similar MOOCs and makes it easier to compare MOOCs with similar characteristics. Some papers share valuable insights. For example, \citet{source20} mention that it is important to take the schedule of students into account when organizing a MOOC. They mention the example of exams that take place in a certain period, which might hinder participation in a MOOC during that exam period. Such considerations or experiences are important to share, as they might not be obvious to others but are valuable to avoid making mistakes that others have already experienced or overcome. 

Seven of the 13 MOOCs (54\%) in this review mention the special role of the class teacher in the implementation of a MOOC in the classroom, mainly to facilitate students to follow the MOOC or guide classroom activities. This seems to have a positive effect on the completion rate of the MOOC. Of the seven papers that mention an above-average completion rate, four describe an experiment with a MOOC in which a classroom teacher plays a role, and two other papers are on MOOCs that can be considered blended because classroom activities are part of the MOOC program. Five papers mention similar or worse completion rates compared to existing MOOC research. The MOOCs in these papers are online MOOCs that are not specifically designed to be used in the classroom but more for individual use. Partly, this is an effect that is expected; when a MOOC is integrated into the classroom dropping out affects school results, which is avoided by many of the stakeholders. Overall, the results show that it is very important to consider the context in which a K-12 CS MOOC is used. The review of K-12 MOOCs by \citet{source3} draws the same conclusion.

Our results show that K-12 CS MOOC research focuses on various aspects related to MOOCs. The aspects are distributed almost equally over the papers. Most research aims to evaluate student behaviour (11 papers, 58\%), followed by learning outcomes (10 papers, 53\%) and student perceptions (9 papers, 47\%), and other goals such as teacher perceptions (6 papers, 32\%). There are no standardised research methods within these categories. Sometimes existing questionnaires are used or adapted. Often, surveys are created for specific research questions, and surveys are not always shared. This makes it harder for other researchers to reuse research methods and put results into context. Using more standardised forms of evaluation can help to get more generalisable results, which can help to create best practices that can be reused within the whole field. 

This review covers the last decade (2014-2024). In more recent years, there were some periods in which education had to move to remote or hybrid settings due to the COVID-19 pandemic. This shift may have had an impact on the development of MOOCs and potentially also on the way that students evaluate MOOCs, because students are more used to remote learning and because a classroom embedding, which has been shown to be important, is missing. However, within the set of papers we reviews, only one study mentioned COVID-19. Maybe papers covering this period are still unpublished.

Computer science is generally a male-dominated field. It is interesting to study whether or not MOOCs attract both genders or whether this inequality can be found in MOOCs too. Fourteen papers mention the gender ratio in their research and only four papers report an approximately equal gender ratio, the other papers again describe experiments in which many more males than females participate. It is unclear whether these gender ratios are also present in computer science courses that do not use MOOCs, or whether this is different for MOOC participation. It would be interesting to further study this, to increase the gender equality in the CS field and to find out if MOOCs can play a role in this development.
Besides attracting more female students, MOOCs might also be adapted such that they better support other underrepresented groups. Only one paper studies how their MOOC is perceived by students with learning difficulties \cite{source27}. 

The findings on (the lack of) diversity are in line with a literature review performed by \citet{van2021participates}. They find that children from the US, boys, and children without previous computer science experience are over-represented in K-12 CS education research. However, often papers do not (consistently) report on the demographics of their participants. Taking divergent groups into account in the design of a MOOC, or designing explicitly for such groups, might make CS more accessible to a wider variety of students \cite{felienne}. Our review indicates that this is lacking in current studies.

\subsection{Threats to validity}
One of the limitations of this review is that we only used academic publications. There are probably MOOCs on CS used in secondary education that are not reported in the scientific literature. In future research, it might be interesting to extend the scope to MOOCs that have not been investigated in academic publications, for example by asking high school CS teachers in multiple countries if they are using any MOOCs in their education. 

The papers we have reviewed use very diverse description methods for their MOOCs, and for the research performed around the MOOCs. Our summary of the results on different aspects might therefore not give a complete overview of the actual results of the studies. For example, we try to summarize the didactical foundations mentioned for the MOOCs. But sometimes this is not (clearly) described in papers, so we might miss some information in this overview. Moreover, one of our findings is that the underlying technologies of MOOCs are often not covered in the papers. It could however be that this is explained in separate papers, which were not included in this review because of our inclusion criteria. For example, student models and mastery algoritms are used in various elearning environments to create an adaptive learning experience. This might be useful for CS MOOCs as well. 

Our review focused on K-12 CS MOOCs, however some included older participants as well. For publicly available MOOCs, it is inevitable that also people from outside the intended user group join the MOOC. 

\section{Conclusion}
\label{sec:conclusions}
In this review, we analyze the current state of the art of K-12 CS MOOCs. We expected that most MOOCs would be specifically designed for K-12 education, since this is such a specific target group. Although we do find many MOOCs for this specific target group or that are based on K-12 curricula, we also find MOOCs that are based on university courses. The review shows that most MOOCs last relatively short, and focus on a specific part of the curriculum. Although MOOCs are typically designed for individual use, this review shows that integrating K-12 MOOCs into a classroom setting in which digital learning and classroom activities are combined leads to higher completion rates. This is in accordance with the findings of \citet{source3}. In such MOOCs, a teacher plays another role: a facilitator, classroom manager, or a coach. For such roles, it is often not essential to have content knowledge. In our review, some papers mention remote support for content questions and grading, with a teacher motivating and supervising the students in the classroom. When used in such a context, MOOCs can be a valuable tool to reduce challenges such as a lack of CS teachers or tools, to integrate CS or CT in the educational portfolio of teachers. However, since MOOCs often only target only a small part of CS, the field needs to mature more to offer a complete K-12 CS curriculum. Moreover, since most MOOCs have a specific focus and we only have a small set of papers, we cannot pinpoint shared factors among the MOOCs in terms of content or design choices beyond the role in the classroom.

It would be helpful if there were more generalizable methods to describe designs of and evaluate MOOCs so that results can be reproduced or generalized more easily. General design guidelines and MOOC evaluations would help new researchers and educational developers to start building upon existing knowledge. In the current situation it is often difficult to determine which design choices are made and hard to compare results between different studies. Although there are existing taxonomies to describe MOOCs, the labels from those taxonomies are rarely used. 

Future research could focus more on the technical and adaptive possibilities that are available for MOOCs. For example, focusing on tracing the knowledge of students and personalising their experience accordingly \cite{Brusilovsky2007}.  
Until now research seems to mainly focus on the design of content and learning materials. In addition, the way that learning materials are delivered, and the way that students are supported by the MOOC itself should be explored more. This can also contribute to the inclusivity of the CS field.

In conclusion, MOOCs can make CS more accessible for K-12 students in different ways: by reducing the challenges that schools are facing when they want to offer CS, or by making them more accessible to a wide variety of students. For a CS MOOC to be successful in engaging students, the MOOC must be mixed with classroom activities. This often implies that the role of the teacher changes, which requires adaptation and sometimes teacher training. We expect that future research will help developing MOOCs that can support schools and teachers to make CS more accessible to everyone.

\section*{Funding}

 This project was funded by EFRO and REACT EU, project number KVW-368 and NOLAI project number 2307.

 \begin{figure}[!h]
  \centering
   \includegraphics[width=5.5cm]{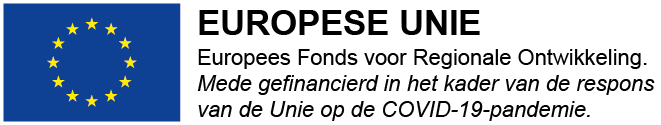}

 \end{figure}

%Bibliography
\bibliographystyle{abbrvnat}  
\bibliography{references}  

\begin{appendix}
\section{Data collection details}
    \label{app:query}

    \begin{table}
\caption{Details on queries used per search engine}
{\begin{tabular}{lp{0.5\textwidth}p{0.2\textwidth}c} \toprule
 Search engine & Query & Additional filters & Results \\ \midrule
 
 Scopus & TITLE-ABS-KEY (("MOOC" OR "MOOCs" OR "Massive Open Online Course" OR "Massive Open Online Courses") AND ("K12" OR "K-12" OR "secondary education" OR "compulsory education" OR "high school" OR "middle school") AND ("computer science" OR "CS" OR "computational thinking" OR "CT" OR "computing" OR "informatics" OR "digital literacy")) AND PUBYEAR $>$ 2013 AND PUBYEAR $<$ 2024 AND (LIMIT-TO (PUBSTAGE ,"final")) AND (LIMIT-TO (DOCTYPE ,"ar") OR LIMIT-TO (DOCTYPE ,"cp")) AND (LIMIT-TO (LANGUAGE ,"English")) &  & 55\\

ERIC & ("MOOC" OR "MOOCs" OR "Massive Open Online Course" OR "Massive Open Online Courses") AND ("K12" OR "K-12" OR "secondary education" OR "compulsory education" OR "high school" OR "middle school") AND ("computer science" OR "CS" OR "computational thinking" OR "CT" OR "computing" OR "informatics" OR "digital literacy") & peer-reviewed only, manual filter on type of publication & 10 \\

IEEE Xplore & ("MOOC" OR "MOOCs" OR "Massive Open Online Course" OR "Massive Open Online Courses") AND ("K12" OR "K-12" OR "secondary education" OR "compulsory education" OR "high school" OR "middle school") AND ("computer science" OR "CS" OR "computational thinking" OR "CT" OR "computing" OR "informatics" OR "digital literacy") & conferences, journals  & 65\\

LearnTechLib & ( "MOOC" OR "MOOCs" OR "Massive Open Online Course" OR "Massive Open Online Courses" ) AND ( "K12" OR "K-12" OR "secondary education" OR "compulsory education") AND ( "computer science" OR "CS" OR "computational thinking" OR "CT" OR "computing" OR "informatics" OR "digital literacy" ) & 2014-2024 & 7 \\

Web of Science & ALL=(("MOOC" OR "MOOCs" OR "Massive Open Online Course" OR "Massive Open Online Courses") AND ("K12" OR "K-12" OR "secondary education" OR "compulsory education" OR "high school" OR "middle school") AND ("computer science" OR "CS" OR "computational thinking" OR "CT" OR "computing" OR "informatics" OR "digital literacy")) & 2014-2024, article & 14 \\

ACM Digital Library & "query": { (Title:( "MOOC" OR "MOOCs" OR "Massive Open Online Course" OR "Massive Open Online Courses" ) OR Abstract:( "MOOC" OR "MOOCs" OR "Massive Open Online Course" OR "Massive Open Online Courses" ) OR Keyword:( "MOOC" OR "MOOCs" OR "Massive Open Online Course" OR "Massive Open Online Courses" )) AND (Title: ( "K12" OR "K-12" OR "secondary education" OR "compulsory education" OR "high school" OR "middle school" ) OR Abstract:( "K12" OR "K-12" OR "secondary education" OR "compulsory education" OR "high school" OR "middle school" ) OR Keyword:( "K12" OR "K-12" OR "secondary education" OR "compulsory education" OR "high school" OR "middle school" )) AND Title:(( "computer science" OR "CS" OR "computational thinking" OR "CT" OR "computing" OR "informatics" OR "digital literacy" ) OR Abstract:( "computer science" OR "CS" OR "computational thinking" OR "CT" OR "computing" OR "informatics" OR "digital literacy" ) OR keyword:( "computer science" OR "CS" OR "computational thinking" OR "CT" OR "computing" OR "informatics" OR "digital literacy" )) }
"filter": { Article Type: Research Article, E-Publication Date: (01/01/2014 TO 05/31/2024) } &
Searched The ACM Guide to Computing Literature & 11 \\

  \bottomrule
\end{tabular}}
\label{tab:queries}
\end{table}

\end{appendix}

\end{document}